\documentclass[letterpaper]{article} 
\usepackage{aaai24}  
\usepackage{times}  
\usepackage{helvet}  
\usepackage{courier}  
\usepackage[hyphens]{url}  
\usepackage{graphicx} 
\usepackage{natbib}  
\usepackage{caption} 
\usepackage{algorithm}
\usepackage{algorithmic}
\usepackage{amsmath}
\usepackage{amssymb}
\usepackage{diagbox}

\usepackage{newfloat}
\usepackage{listings}
\DeclareCaptionStyle{ruled}{labelfont=normalfont,labelsep=colon,strut=off} 
\floatstyle{ruled}
\newfloat{listing}{tb}{lst}{}
\floatname{listing}{Listing}
\usepackage{hyperref} 

\title{Operationalizing content moderation ``accuracy'' in the Digital Services Act}
\author {
    Johnny Tian-Zheng Wei\textsuperscript{\rm 1},
    Frederike Zufall\textsuperscript{\rm 2,3},
    Robin Jia\textsuperscript{\rm 1}
}
\affiliations {
    \textsuperscript{\rm 1}University of Southern California\\
    \textsuperscript{\rm 2}Karlsruhe Institute of Technology\\
    \textsuperscript{\rm 3}Waseda University\\
    jtwei@usc.edu, zufall@kit.edu, robinjia@usc.edu
}

\usepackage{bibentry}

\begin{document}

\maketitle

\begin{abstract}
The Digital Services Act\footnote{Regulation (EU) 2022/2065 of the European Parliament and of the Council of 19 October 2022 on a Single Market For Digital Services and amending Directive 2000/31/EC (DSA).}, recently adopted by the EU, requires social media platforms to report the ``accuracy'' of their automated content moderation systems. The colloquial term is vague, or open-textured---the literal accuracy (number of correct predictions divided by the total) is not suitable for problems with large class imbalance, and the ground truth and dataset to measure accuracy against is unspecified. Without further specification, the regulatory requirement allows for deficient reporting. In this interdisciplinary work, we operationalize ``accuracy'' reporting by refining legal concepts and relating them to technical implementation. We start by elucidating the legislative purpose of the Act to legally justify an interpretation of ``accuracy'' as precision and recall. These metrics remain informative in class imbalanced settings, and reflect the proportional balancing of Fundamental Rights of the EU Charter. We then focus on the estimation of recall, as its naive estimation can incur extremely high annotation costs and disproportionately interfere with the platform's right to conduct business. Through a simulation study, we show that recall can be efficiently estimated using stratified sampling with trained classifiers, and provide concrete recommendations for its application. Finally, we present a case study of recall reporting for a subset of Reddit under the Act. Based on the language in the Act, we identify a number of ways recall could be reported due to underspecification. We report on one possibility using our improved estimator, and discuss the implications and areas for further legal clarification.
\end{abstract}

\section{Introduction}

Many concerns have been raised about the role of social media in spreading harmful content such as online hate \cite{siegel2020online}, and  
these concerns persist despite the fact that
social media companies voluntarily moderate their content \cite{gillespie2018custodians}. On the other hand, content moderation raises the question of whether companies are interfering with users' freedom of speech.
Research continues to demonstrate that voluntary moderation is problematic, 
and these issues will remain unresolved so long as social media companies have conflicting interests in prioritizing platform growth over providing safe online spaces \cite{pelley_2021}. 

In light of these conflicting interests, legislators have increasingly sought to enact legal requirements for content moderation. 
Regulating content moderation is a balancing act between conflicting rights: on one hand there is a need to protect the right to the freedom of speech, and on the other to curtail societal harms \cite{douek_governing_2021}. This balance must be made all while considering the technical feasibility of such legal requirements, so as to respect companies' right to conduct business. Transparency reporting obligations, which require companies to release reports about their content moderation systems, are the first and necessary step in the regulatory landscape. Like any legal instrument, transparency requirements must respect the rights of all entities involved, and be defined with care to ensure their relevance.



The Digital Services Act, recently adopted by the EU, is a milestone legal instrument addressing content moderation.
It includes a transparency obligation for social media platforms to report the ``accuracy'' of their content moderation systems. \textbf{However, the colloquial ``accuracy'' is underspecified, and the regulatory requirement, as it stands, allows for deficient reporting.} This is the departure point of our work: we operationalize ``accuracy'' by refining legal concepts and relating them to technical implementation. Our work provides an interpretation of content moderation ``accuracy'' in the Digital Services Act as precision and recall, which we show is both legally and technically sound. By providing this interpretation, we highlight areas for further legal clarification to ensure meaningful reporting. The argumentation here can be useful for future regulatory guidelines or legislative amendment.

Our interdisciplinary contribution can be viewed from two perspectives. From a legal perspective, we identify the sources of underspecification in the ``accuracy'' reporting requirement and show that precision and recall are appropriate instantiations which reflect the proportional balancing of Fundamental Rights of the EU Charter. From a technical perspective, we formulate the reporting of recall as unbiased statistical estimation and show how such estimation can be made efficient with stratified sampling. With the improved estimator, we demonstrate the feasibility of recall reporting in a case study on Reddit.
Our paper contains three parts:

\begin{enumerate}
    \item \S \ref{sec:dsa} starts with elucidating the legislative purpose of the Act to legally justify an interpretation of ``accuracy'' as precision and recall. These metrics remain informative in class imbalanced settings and reflect the proportional balancing of the Fundamental Rights of the EU Charter.
    
    \item \S \ref{sec:statistically_estimating_law} formulates the reporting of recall as an unbiased statistical estimation problem. Since its naive estimation can pose an disproportionate burden for companies, we propose the use of stratified sampling. In \S \ref{sec:simulated_results} we  derive practical recommendations for applying stratified sampling through a simulation study.
    
    \item \S \ref{sec:reddit_prevalence} presents a case study of a recall reporting requirement for Reddit under the Digital Services Act. Based on the language in the Act, we identify a number of ways recall could be reported due to underspecification. We choose one possibility to report, and discuss the implications and areas for legal clarification.
\end{enumerate}

\section{Background and related work}
\label{sec:related_work}

\paragraph{Count-based metrics} At the time of writing, many social media companies voluntarily report basic content moderation statistics. Even without direct regulation, companies often provide transparency reports to build public trust in their platform.\footnote{\url{https://integrityinstitute.org/news/institute-news/integrity-institute-releases-overview-of-online-social-platform-transparency}} 
For instance, Twitter reports a few count-based metrics such as the number of posts that Twitter took moderation action against in a given reporting period.\footnote{\url{https://transparency.twitter.com/}} 
However, these basic statistics conflate multiple phenomena. When the number of takedowns of hate speech increases, it is not immediately clear whether it is due to an increase in hate speech, an increase in moderation, or a change in their content moderation policy.

\paragraph{Prevalence metrics} To the best of our knowledge, only Facebook, Youtube, and Snap publicly provide prevalence metrics. Facebook reports the prevalence of different categories of content violations over time (e.g. 2 in 1000 posts contained hate speech)\footnote{\url{https://transparency.meta.com/policies/improving/prevalence-metric/}}. As prevalence is a normalized measure (the number of content violations divided by the total amount of content on the platform) it is suitable to track increases in content violations. While their exact methods to compute prevalence are unclear, they note that ``we calculate this metric by selecting a sample of content seen on Facebook and then labeling how much of it shouldn’t be there.''\footnote{\url{https://about.fb.com/news/2019/05/measuring-prevalence/}} Facebook's own internal studies deem prevalence as a major statistic and industry best practice \cite{fb_independent_2019}. Research in social media has long used prevalence in online discussions to study the impact of real world events. For example, \citet{DBLP:conf/icwsm/OConnorBRS10} showed that presidential job approval can be well estimated by the prevalence of tweets mentioning the president which have positive sentiment. 

\paragraph{Recall} There is a direct connection between prevalence and recall. By dividing the number of violations acted upon (i.e. the number of takedowns) against the total number of violations (the total number of posts multiplied by violation prevalence, see \S\ref{sec:setup_recall}), we can calculate the recall of content moderation, or rate at which content violations are identified and acted against. Recall is a measure of moderation ``thoroughness'', and we speculate that these large companies may be computing a moderation recall internally to monitor moderation performance, by comparing prevalence with other statistics, such as number of takedowns \cite{liu_unofficial_2019}.

\paragraph{Metrics required by regulation} Almost all major social media platforms provide transparency reports for German users as specified by Germany's Network Enforcement Act (NetzDG).\footnote{Netzwerkdurchsetzungsgesetz vom 1. September 2017 (BGBl. I S. 3352).} NetzDG requires companies to periodically provide information and statistics related to the content moderation conducted at these companies (requirements which are succeeded by those in the Digital Services Act). Amongst the required information is also the community guidelines by which companies moderate content. These reports demonstrate the potential of regulation to standardize the set of transparency metrics reported on. By relating existing transparency requirements to content moderation recall,  our work highlights how further legal clarification can ensure meaningful reporting. Recall could reflect general moderation effectiveness, and requiring it would encourage companies to periodically conduct studies of their platform design \cite{Harling_Henesy_Simmance_2023}.

\section{The Digital Services Act}
\label{sec:legal_overview}
\label{sec:dsa}

The Digital Services Act (DSA), fully applicable since February 17th, 2024, introduces a comprehensive new framework for content moderation. 
Social media providers with more than 45 million average monthly active recipients in the Union (Art. 33(1) DSA), which currently includes U.S.-established platforms Twitter, Instagram, Youtube, and Facebook, are classified as ``very large online platforms'' (VLOPs) and will have to comply with the full set of requirements set forth in the Act.\footnote{\url{https://ec.europa.eu/commission/presscorner/detail/en/ip_23_2413.}} This work focuses on the transparency reporting obligations on content moderation introduced by Art. 15(1).

\paragraph{Territorial scope} The DSA is a milestone legal instrument targeting content moderation. Hence, our study of the Act may have an impact beyond the EU---a phenomenon termed the ``Brussels effect'' \cite{bradford_brussels_2012}.
For one, nation-states outside of the EU will likely reference the DSA in considering legislation for their own jurisdictions. As companies may already have the infrastructure to satisfy EU requirements, adopting all or some of the DSA poses a low regulatory burden. Moreover, the DSA (like the GDPR) has a broad territorial scope, and requires companies operating in the EU to apply EU standards to all users, regardless of their nationality (Art. 2(1) DSA).

\paragraph{Considerations for future implementation} The Commission has begun to adopt implementing and designating decisions for the DSA.\footnote{\url{https://digital-strategy.ec.europa.eu/en/policies/digital-services-act-package}} In addition, a new European Centre for Algorithmic Transparency has been established to support the Commission in analyzing transparency reports.\footnote{\url{https://algorithmic-transparency.ec.europa.eu/}} The effects of the Regulation will have to be reviewed by 2025 and 2027 (Art. 91 DSA)---allowing for an ongoing debate on its implementation and potential future amendments.

\subsection{Scope of content moderation} \label{sec:content_moderation}


\paragraph{Definition of ``content moderation''} The DSA defines ``content moderation'' as ``activities [\dots] that are aimed, in particular, at detecting, identifying, and addressing'' (Art. 3(t) DSA) the two following classes of content:
\begin{itemize}
    \item \emph{Illegal content.} This is defined as ``any information that [\dots] is not in compliance with Union law or the law of any Member State [\dots], irrespective of the precise subject matter or nature of that law'' by Art. 3(h) DSA. For instance, national law of the EU Member States must contain a minimum standard to incriminate hate speech as specified by an EU Framework Decision.\footnote{Framework  Decision  2008/913/JHA  of  28  November 2008 on combating certain forms and expressions of racism and xenophobia by means of criminal law and national laws transposing it.} Furthermore, the EU Commission has also started an initiative to add ``all forms of hate crime and hate speech, whether because of race, religion, gender or sexuality''\footnote{Communication of 9.12.2021, COM(2021) 777 final. This would allow the Commission to replace the existing Framework Decision by a new Directive further elaborating on a more extensive notion of hate speech incrimination. See also the recently adopted European Parliament Resolution of 18.01.2024, P9\_TA(2024)0044.} to the list of EU crimes in Art. 83(1) TFEU.\footnote{Treaty on the Functioning of the European Union, [2016] OJ C202/1.}
    
    \item \emph{Content that violates terms and conditions.} 
    Another content moderation scenario that the DSA specifically addresses are any activities that are aimed ``at detecting, identifying and addressing [\dots] information incompatible with their terms and conditions'' (Art. 3(t) DSA). 
    This includes the ``community guidelines'' as a part of the terms and conditions. 
\end{itemize}
The DSA acknowledges a wide range of measures to ``address'' this identified content, including ``measures taken that affect the availability, visibility, and accessibility of that illegal content or that information'' (Art. 3(t) DSA).

\paragraph{Relation to Fundamental Rights}
\label{sec:balancing}

The DSA, like any other regulatory instrument in EU law, is bound and to be interpreted and applied in accordance with the EU Charter of Fundamental Rights (CFR)\footnote{Charter of Fundamental Rights of the European Union, OJ C 326, 26.10.2012, p. 391.}.
While they only directly bind EU institutions (Art. 51 CFR), they also affect providers, as they guide the interpretation of provisions which regulate relations between citizens and private entities.\footnote{ECJ, C-360/10 – Sabam, 26.2.2012, para. 52.}
Namely, Recital(3) of the DSA explicitly refers to three protected rights that are relevant here:

\begin{itemize}
    \item \emph{Freedom of expression and information (Art. 11 CFR)} protects online content from censorship and moderating measures. Moderating online posts interferes with this right and requires justification.
   
    \item \emph{The right to non-discrimination (Art. 21 CFR)} forbids any discrimination based on grounds like sex, race, colour, ethnic or social origin and is especially reflected in the EU framework against the expression of hatred \cite{zufall_hate_speech}.\footnote{Usually, national constitutions contain a similar guarantee which is again reflected in national laws, e.g., Criminal Law, that once more constitute "illegal content" in the sense of the DSA.} To prevent discrimination, this right may provide the justification for the interference with freedom of expression.

    \item \emph{Freedom to conduct a business (Art. 16 CFR)} is equally mentioned by the DSA and relevant for our context as any obligation imposed by the DSA to providers is interfering with this right. In addition to balancing out freedom of expression and the purpose of non-discrimination, the freedom to conduct a business also requires proportionality. 

\end{itemize}
    

Balancing out these rights is a task not only performed by the legislator upon drafting regulation like the DSA, but also at the time of its subsequent executive implementation and judicial interpretation (Art. 52(5) CFR). 
Any automated content moderation system subject to the DSA that interferes with these rights requires a legal base in the form of law (Art. 52(1) CFR). 
This requirement reflects the democratic idea that any person who is subject to an interference must have the possibility to influence the origin of that interference through democratic elections. Here, the origin of interference is the content moderation system and the obligation to moderate content. Subsequently enacted laws by the legislation can again then build the legal foundation for limiting and balancing out these fundamental rights.
Therefore, these rights continue to be relevant not only for regulatory implementing decisions, but also at the level of technical implementation.

\subsection{``Accuracy'' as a transparency requirement}
\label{sec:interpreting_accuracy}

Transparency requirements in the DSA not only are important legal instruments for revealing how content moderation systems work, but also for monitoring how they interfere with fundamental rights. 
Transparency requirements may also reflect the degree to which this balancing of rights has been carried out in favor of one right towards another. 
Therefore, transparency not only enables democratic control and judicial review, but also offers insight into the extent democratic legitimacy is carried through, to the final step of technical implementation.

\paragraph{Transparency requirements in the Act}
The DSA requires that Terms and Conditions must contain information on any policies, procedures, measures and tools used for the purpose of content moderation, including algorithmic decision-making and human review (Art. 14(1) DSA).
Even further, providers are obliged to publish comprehensive reports at least once a year---in the case of VLOPs, every six months---on any content moderation they engaged in during the relevant period (Art. 15(1), Art. 42(1),(2) DSA). 
In all these cases, Art. 15(1)(e) DSA requires providers to report:
\begin{quote}
\small
``any use made of automated means for the purpose of content moderation, including a qualitative description, a specification of the precise purposes, \textbf{indicators of the accuracy and the possible rate of error of the automated means} used in fulfilling those purposes, and any safeguards applied.''
\end{quote}
In addition, VLOPs must also conduct mandatory risk assessments that include their content moderation systems (Art. 34 DSA). 

\paragraph{Troublingly, the DSA leaves the term ``accuracy'' underspecified.} Art. 15(1)(e) DSA only asks for ``indicators of the accuracy and the possible rate of error".\footnote{The current preparatory document for a draft on a Commission Implementing Regulation laying down templates concerning the transparency reporting obligations (Ares(2023)8428591) does not provide further guidance, and only asks for the reporting of an ``accuracy rate'' and ``error rate'' as quantitative information in its templates.} From a technical perspective, the term ``accuracy'' raises a few questions:  if we are estimating accuracy in its literal sense (correct number of predictions divided by the total), to which ground truth are we measuring against? In addition, accuracy is not suitable for problems with large class imbalance, so what metrics would be more appropriate? We identify ``accuracy'' as an \textit{open-textured} term, which needs interpretation in novel contexts. As it stands, ``accuracy'' reporting required by the DSA will exhibit inconsistency, and social media providers may each make their own decisions regarding test data and evaluation metrics. They could choose to report the metric most beneficial for themselves, which would undercut the purpose of the regulation. By providing a legally and technically sound interpretation of ``accuracy'', our work highlights areas for further legal clarification to avoid an enforcement deficit.

\paragraph{What does ``accuracy'' refer to?} Interpreting the notion of ``accuracy'' beyond its literal wording needs to be in line with the legislative purpose of the Act and requires justification. Future legal adaptations of our analysis depends on this legal justification.
We attempt to elucidate and define the notion of ``accuracy'' in Art. 15(1)(e) of the DSA using a method of systematic interpretation in EU law: inter-instrumental interpretation \cite{grundmann_2011}. By referencing the definitions and systematic understanding from another instrument of EU law, we can make an attempt to apply it to ``accuracy.''
We refer to the upcoming Artificial Intelligence Act (AI Act),\footnote{Compare the latest consolidated version adopted by the European Parliament: P9\_TA(2024)0138.} a parallel development in EU law related to specific high-risk AI systems.
Similar to the DSA, the AI Act applies to U.S.-based providers if they are ``placing on the market or putting into service AI systems in the Union'' (Art. 2(1)(a) AI Act). 
While the AI Act is specifically targeted towards high-risk AI systems, it outlines a general regulatory approach to automated classification systems. 

With respect to ``accuracy'', Art. 15(1) of the AI Act requires ``(high-risk) AI systems to be designed and developed in such a way that they achieve, in the light of their intended purpose, an appropriate level of \textit{accuracy} [\dots] and perform consistently in those respects throughout their lifecycle.''
In this context, the AI Act (Art. 15(2)) calls upon the Commission to encourage ``the development of benchmarks and measurement methodologies'' for, inter alia, ``the appropriate levels of accuracy''.
However, similarly to the DSA, the Act only mentions ``accuracy'' and not other metrics such as precision or recall. Upon further inspection, however, the legislative process reveals that the Act was inspired by the policy guidelines of the High Level Expert Group on AI \cite{HLEG_ALTAI}. In these guidelines, the use of accuracy is intended to reflect the ``trustworthiness'' of an AI system, but a complementary footnote stated that ``accuracy  is  only  one  performance  metric  and  it  might  not  be  the  most  appropriate  depending  on  the  application,'' and that the F1 score, false positives, and false negatives may also be used. Hence, any interpretation of ``accuracy'' beyond its literal wording in the DSA, should use an ``appropriate metric depending on the application".

\subsection{Metrics and their legal implications} \label{sec:metrics}

Having opened up ``accuracy'' to the possibility of broader interpretation, we consider its operationalization with basic accuracy metrics, and discuss their legal implications.

\paragraph{TP, FP, TN, and FN} TP, FP, TN, and FN are the true positives, false positives, true negatives, and false negatives, respectively. This section discusses content moderation in terms of classifying illegal content (which is a perspective the DSA takes, see \S\ref{sec:content_moderation}). In this context, TP and TN represent correctly identified illegal and free speech. Each FP is an individual violation to the freedom of expression (by misclassifying and acting against free speech), while each FN is a failure to identify and act against discriminatory content. 

TP, FP, TN, and FN are primitives from which classification metrics are derived. Therefore, the reporting of TP, FP, TN, and FN, would be sufficient to calculate accuracy, precision, recall, F1, or any other metric. As FN and FP directly correspond to the interference of Fundamental Rights (listed in \S\ref{sec:content_moderation}), we encourage regulators to reason about precision and recall, which are their normalized measures. Metrics such as accuracy or F1 (the harmonic mean of the precision and recall) can serve as useful summary statistics. However, only reporting summary statistics will obscure whether platforms are over- or undermoderating. 

\paragraph{Accuracy}

This metric is by far mentioned the most in EU law, and defined as:
$\text{Accuracy} = \frac{TP+TN}{TP+TN+FP+FN}$which is the number of correct predictions divided by the total. Using accuracy as a metric for content moderation is problematic because of label-imbalance in the content moderation setting: in a huge set of content like social media posts, the illegal posts are usually far outnumbered by the posts that do not violate specific laws. Accuracy can be highly inflated if the test data is label-imbalanced, as a classifier that only outputs the majority label will achieve high accuracy.  In \S \ref{sec:keyword_filter}, we show that even a simple classifier can achieve an accuracy of 95\% due to label imbalance in the data set. Alternatively, precision and recall are separate accuracies of the predicted positives and true positives, respectively, which remain informative in label imbalanced settings.

\paragraph{Precision} 

Precision measures the fraction of examples positively predicted (e.g., identified by a content moderation system for removal) that are truly positive (e.g., illegal content, depending on the ground truth):

\begin{equation}
    \text{Precision} = \frac{TP}{TP+FP}.
\end{equation}
As each FP is an individual violation to the freedom of expression (by misclassifying and acting against free speech), precision is a probabilistic reflection of the moderation system's respect to the right of freedom of expression. A low precision would indicate that much content which is not actually illegal is being positively identified and acted against. If a social media platform exceeds their legal obligation of content moderation, overmoderation would interfere with the right to freedom of expression. Content moderation is only justified if it does not present a disproportionate interference to this right, which is measured by precision.

\paragraph{Recall} 
Finally, recall measures the fraction of truly positive content (e.g., illegal posts) that was predicted positive (e.g. identified by the moderation system for removal): 
\begin{equation} \label{eq:recall}
    \text{Recall} = \frac{TP}{TP+FN}.
\end{equation}
As each FN is an individual failure to identify and act against discriminatory content, recall is a probabilistic reflection of the moderation system's respect to the right to non-discrimination. A low recall would indicate much illegal content is still visible on the platform. In this sense, recall well satisfies the purpose for which the DSA imposes a reporting obligation: to ensure compliance with EU or national laws against illegal content.

\paragraph{The precision-recall tradeoff} The DSA has to balance the freedom of expression with the right to non-discrimination, and such a balance would be reflected mathematically by the precision-recall tradeoff.\footnote{\url{https://scikit-learn.org/stable/auto_examples/model_selection/plot_precision_recall.html}} A probabilistic classifier (i.e. one that predicts the probability of whether a given piece of content is illegal, and takes an action against it, such as removal, if the score is greater than a decision threshold) can trade off precision for recall and vica versa by tuning its decision threshold. By increasing the threshold, the classifier is more conservative in its positive predictions, which increases the precision at the cost of recall. Likewise, precision can be traded off for recall by decreasing the threshold. The decision threshold is an implicit but intentional setting set by the platforms that deploy these classifiers, and requiring the reporting of precision and recall would make clear the effects of this implicit threshold \cite{douek2021governing}. Regulators should recognize that, while improving both precision and recall simultaneously is non-trivial, trading them off is a technical option to strike the appropriate balance between fundamental rights.

\section{Statistically estimating legal requirements} \label{sec:statistically_estimating_law}

Given that interpreting content moderation ``accuracy'' as precision and recall is legally sensible, we now discuss the technical implementation of precision and recall. This section starts with an example of statistical estimation intended for an interdisciplinary audience.

As an example, suppose a social media platform deploys an automated content moderation system that deletes content which violates their community guidelines. We may want to estimate the prevalence (or proportion) of content violations which remain visible on the platform. This prevalence would be the false negative rate of the deployed moderation system, and can be used to calculate the recall (more details in \S\ref{sec:setup_recall}). We outline three components of statistical estimation here:

\paragraph{Expectation}  Statistical estimation begins with positing the ideal quantity we are estimating. Ideally, an annotator could label every single piece of visible content on the platform to determine whether it is violating or not. We could then directly calculate the false negative rate, which we denote as $p$ (where $0\leq p\leq 1$). 
Since $p$ is derived from an exhaustive average, $p$ is called an \textit{expectation} in statistical terms. 
However, social media platforms accumulate millions of pieces of content a day, so exhaustive annotation would not be feasible. Therefore, the true $p$ will be unknown.

\paragraph{Sampling} The only practical alternative is to estimate the false negative rate $p$ by \textit{sampling} some visible content and then annotating this sample. The prevalence of content in the sample, which we denote as $\hat{p}$, can be used as an estimate for $p$. The number of samples directly relates to the implementation cost: each sample needs to be labelled by an annotator.

\paragraph{Standard error} A better estimate $\hat{p}$ is one that is close to $p$, on average. The average distance between $\hat{p}$ and $p$ is a \textit{standard error} in statistical terms. If the estimate $\hat{p}$ is unbiased, more samples results in smaller standard error.  Estimation error is then a function of these two components:
\begin{itemize}
    \item \textbf{Unbiasedness} For legal reporting, we should provide a \textit{statistically unbiased} estimate $\hat{p}$ which, on average, reflects the false negative rate $p$. When estimating $\hat{p}$, two details ensure statistical unbiasedness. The first is using a ``representative'' sampling method like random sampling. Stratified sampling is also unbiased by emulating a ``representative'' sample (see \S\ref{sec:stratified_sampling}). Second, human ground truth annotations are required. Solely relying on classifiers introduces bias because they make errors relative to the ground truth \cite{gorwa_algorithmic_2020}. 
    
    \item \textbf{Number of samples (cost)} An unduly burdensome reporting requirement would interfere disproportionately with the right to conduct business (Art. 16 CFR). Thus, our statistical analyses focuses on cost: \textit{how many} annotations are needed to provide a reasonably good estimate? In general, a larger sample yields better estimates. On the other hand, as $p$ approaches 0, it requires more samples to accurately estimate e.g. if $p=0.02$, 10 samples would likely yield estimates $\hat{p}=0.0$ or $\hat{p}=0.1$, where the sample contains 0 or 1 pieces of violating content. More samples would be needed to refine the granularity of $\hat{p}$.
\end{itemize}

\subsection{Estimating precision}

Auditing content moderation precision is qualitatively difficult but statistically simple, and so is not the focus of our work. In short, precision can be well estimated by annotating a random sample of positively predicted content. Appendix \ref{appendix:estimating_precision} contains a discussion on estimating precision.

\subsection{Estimating recall}
\label{sec:setup_recall}

Measuring recall poses a challenging statistical problem, which is the focus of our work. To compute recall, both TP and FN are needed. TP can be computed from the precision (the positives multiplied by the precision), but estimating FN requires looking through the predicted negatives of the content moderation system (i.e. visible content on the platform during the last reporting period) to determine the number of false negatives. Similarly, recall is challenging to measure for tasks like information retrieval, as the false negatives are the relevant webpages missed by the retrieval system across the web \cite{aslam_statistical_2006}. 

Typically, the prevalence of misclassified violations amongst all content is rare, e.g. 1 in 1000, so estimating the false negative rate requires a large number of samples to provide a reasonable estimate. Appendix \ref{appendix:formulation} contains the statistical formulation on estimating the false negative rate $p$. The simplest way to estimate $p$ is to apply the random sampling estimator, where a random sample of negatives is collected and annotated to compute $\hat{p}$ as an estimate for $p$. When estimating a small value like $p$, we want the estimation error to be small, relative to $p$. In other words, we are estimating the rare probability $p$ within some percentage accuracy. In statistical terms, we want to ensure the coefficient of variance $\textsc{CV}_{\hat{p}} = \textsc{SE}_{\hat{p}} / \hat{p}$ is low i.e. $\textsc{CV}_{\hat{p}} \leq 20\% $, where $\textsc{SE}_{\hat{p}}$ is the standard error of estimate $\hat{p}$ \cite{precision_of_estimates}.

\subsection{Stratified sampling}
\label{sec:stratified_sampling}

To reduce the burden of recall reporting, we propose using stratified sampling instead of random sampling. Stratified sampling can provide a better estimate with the same number of samples \cite{mcbook}. Crucially, it is \textit{unbiased} and \textit{more efficient} than a random sampling estimator. Appendix \ref{appendix:ss_notation} contains notation for stratified sampling and only a high level discussion is given here. The procedure is as follows: a learned classifier is first used to partition e.g. the visible comments into bins. Then, the estimator is computed by annotating a random sample within each stratum, and combining the prevalences across stratum in an unbiased manner.

\paragraph{Intuition} A good stratification can significantly increase the efficiency of the stratified sampling estimator over random sampling. If some strata have a very low or very high prevalence of violating comments, few samples are needed in those strata to estimate the prevalences well. This allows us to allocate our annotation effort on estimating strata with medium prevalences (which have higher variance). Stratified sampling is unbiased even if the stratification classifier has undesirable properties such as low accuracy or low fairness \cite{DBLP:conf/www/BorkanDSTV19}, as a human is the final decision maker. A suboptimal classifier may cause the estimator to be less efficient, but it cannot introduce statistical bias.

\subsection{Binning and allocation}
\label{sec:binning}
The variance of stratified sampling depends on two choices: binning (i.e., how to partition the examples) and allocation (i.e., how many examples to annotate in each stratum).
We now provide technical details for each choice.

\paragraph{Binning} An effective way to create strata is with a binning classifier trained to predict the probability that a given piece of content violates community guidelines.
With the predictions scores from the binning classifier, content can be binned in a few ways by their score.  The binning classifiers gives each comment a score from $[0, 1]$.  The number of bins $L$ is a hyperparameter, which we investigate in \S \ref{sec:simulated_results}.
\textbf{(1)} \emph{Equal width binning}. Equal-width binning splits this interval into $L$ equal width intervals each corresponding to a bin. Sampling across bins thus samples across the entire scoring range of the classifier.
\textbf{(2)} \emph{Quantile binning}. Once the comments are scored, quantile binning splits the scoring range into equal sized quantiles, where each bin will contain the same number of points. The range of scores within each quantile will depend on the scoring distribution of the classifier.
\textbf{(3)} \emph{Oracle binning}. This binning is not possible in practice as it requires the labels of all the data in the pool, but we include it as an oracle method. Using a recursive search procedure and the labels of all the training data, we can search for a good binning. At each recursive step, a bin is split into two at a point where the resulting variance is minimized. We recursively break down the binning classifier's scoring interval into a number of bins that is a power of two.

\paragraph{Allocation} Once the bins are partitioned, an allocation of how to sample from each strata (i.e., a choice of the $n_h$'s) can be made. 
\textbf{(1)} \emph{Equal allocation.} This baseline allocation samples from each bin equally. 
\textbf{(2)} \emph{Optimal allocation.} 
For a given stratification, an optimal sample allocation minimizes the variance of the estimator. The solution is given in Appendix \ref{appendix:optimal_alloc}.
This allocation is an oracle method: it cannot be run in practice, as it requires the standard deviations within each stratum, which are not known beforehand. We include this allocation method for analysis purposes. 
\textbf{(3)} \emph{Pilot allocation.} To approximate the optimal allocation, we adopt a simplified version of \citet{DBLP:conf/cikm/BennettC10}. We annotate a fixed number of pilot samples within each stratum, which we use to estimate the standard deviation per stratum. We then approximate the optimal allocation using these standard deviations. If the size of the pilot sample within the stratum exceeds the optimal allocation for that stratum, no additional samples are collected for that stratum. If the optimal allocation is more than the pilot sample, the additional samples are collected. 
We use psuedocounts (adding 1 positive and 1 negative example) to ensure each stratum is allocated samples. 

\subsection{Analyzing the burden of estimation}\label{sec:burden_estimation}

The statistical analyses here are focused on cost i.e. how many annotations are needed to provide a good estimate? The number of annotations needed to achieve a fixed estimation error can be calculated analytically for both random and stratified sampling estimators. A few technical details are provided here and we defer the rest to the appendix. Appendix \ref{appendix:rs_power} and \ref{appendix:cost_derivations} contain closed form solutions for the cost of random and stratified sampling, respectively. To give some intuition, the cost of estimating different prevalences with random sampling are given in Appendix \ref{appendix:rs_power}.

We consider the goal to report $\hat{p}$ to within $20\%$ of $p$, with 95\% probability. In statistical terms, we calculate the number of samples required for the 95\% confidence interval of $\hat{p}$ to be a fixed width relative to $p$.\footnote{Once we have a statistical estimate $\hat{p}$ for the false negative rate $p$, the confidence intervals on $\hat{p}$ can propagate to a confidence interval on recall. In the case where we know the exact number of true positives $TP$, the confidence intervals on $p$ can directly be translated to confidence intervals on $FN$, and then to recall, by plugging in the upper and lower estimates of $p$ into the recall equation. If $TP$ is not a known quantity and also an estimate, confidence intervals can be obtained through bootstrap simulation.} The coefficient of variance is effectively required to be $\textsc{CV}_{req} = 0.2 / z_{.95}$, and in turn the standard error is required to be $\textsc{SE}_{req} = (0.2) \hat{p} / z_{.95}$, where $z_{(1-\alpha)}$ is test statistic of the normal variable. Since $\textsc{SE}_{req}$ is a function of the total number of annotations collected, we can solve for this number for each estimator. 

\section{Efficiently estimating recall for a filter} \label{sec:simulated_results}

In this section, we determine which choices of binning, allocation (described in \S \ref{sec:binning}), and classifier are best for estimating recall with stratified sampling. The efficient and robust estimator derived in this section is used to estimate a content moderation recall for Reddit in \S\ref{sec:reddit_prevalence}, where we will continue to provide legal analysis on areas of underspecification in the ``accuracy'' reporting requirement.

We use CivilComments \cite{DBLP:conf/www/BorkanDSTV19} as our testbed and emulate a content moderation system by constructing a keyword-based comment filter.
While Civilcomments is annotated for toxicity and is not an operationalized legal definition or community guideline, this setting provides a large dataset to study recall estimation on a related task.
Similar to pool-based active learning \cite{DBLP:series/synthesis/2012Settles}, we assume in our simulated experiments that the dataset is initially unlabeled, and we ``annotate'' whether an example is a false negative by revealing the label from the dataset.

\subsection{A simple keyword filter}
\label{sec:keyword_filter}
To emulate an automated content moderation system, we construct a keyword-based filter which removes comments in the Civilcomments dataset (see \S \ref{sec:keyword_exp_setup}). We choose the list of keywords using the top 10 positive features from an unigram logistic regression trained to predict comment toxicity.

In total, the dataset contains about 100K toxic comments out of 1.7M comments, and the overall prevalence of toxicity is 5.9\%. The filter achieves a precision of 58\%, removing 57K comments with 32K of these actually being toxic. Amongst the unfiltered examples, the prevalence is 4.1\%. The filter would only achieve a recall of 33\%, removing about 33K of the total 100K toxic comments. However, the accuracy would be nearly 95\%.

Note that the high accuracy obscures the fact that only 33\% of the toxic comments were found and removed. In our simulated experiments, the goal is to statistically estimate the recall using as few labeled examples as possible. By estimating the prevalence of toxicity in the unfiltered comments, we can calculate recall. In our experiments, we compare costs when estimating the prevalence of 4.1\% to within 20\%. If prevalence is estimated as such, the confidence intervals on the recall would translate to $[28.9\%, 37.9\%]$.

\subsection{Experimental setup}
\label{sec:keyword_exp_setup}

\paragraph{Dataset} The CivilComments dataset \cite{DBLP:conf/www/BorkanDSTV19} contains comments labeled by their toxicity, from the archive of the Civil Comments platform, a commenting plugin for independent news sites. Each comment is annotated for toxicity by at least 10 annotators, and if half of the annotators label it toxic, the toxicity label is positive. For our study, we use 100K random examples of the training set for training, and the rest of the 1.7M training examples for simulation. This creates a large testing setting which better represents content moderation, where the number of examples to train the model is small but it is used for a large number of predictions. This dataset naturally presents a rare prevalence estimation problem as $5.9\%$ of comments are toxic.

\paragraph{Binning classifier} Roberta \cite{DBLP:journals/corr/roberta} is a transformer-based language model pretrained on a large corpus of English data, which has strong performance for CivilComments after finetuning. We use typical hyperparameters for small datasets \cite{DBLP:conf/iclr/MosbachAK21}, and make two critical adjustments: increasing the batch size to 32 (we hypothesize that since the labels can be noisy, the gradients are unstable), and subsampling the negative examples so that the training data is balanced (the resulting training data has 11k examples).

\subsection{Discussion}
\label{sec:ss_results}

\begin{figure*}
    \centering
    \includegraphics[scale=0.94]{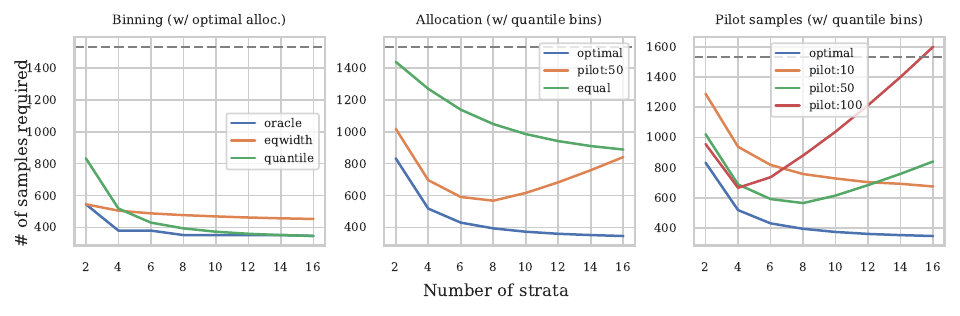}
    \caption{Number of samples required to estimate the prevalence of toxicity (4.1\%) within 20\% in the unfiltered CivilComments, with different number of strata. Binning is based on predicted scores from a finetuned Roberta model. All pilot results are averaged over 30 trials. Dotted line represents random sampling. As the number of strata grows, estimators using optimal allocation require fewer samples. Pilot methods outperform the equal allocation baseline, but when there are too many strata, annotation effort is wasted on the pilot samples.}
    \label{fig:binning_allocation}
\end{figure*}

\paragraph{Results} As shown in Figure \ref{fig:binning_allocation},
using the right techniques and parameters for stratified sampling makes a large difference: the worst choices perform as poorly as the random sampling estimator, while the best choice can save up to half of the annotation effort. 
With optimal allocation, increasing the number of strata decreases variance.
Intuitively, optimal allocation is more effective with more bins because a more precise allocation is given to each strata, according to the strata prevalence. The pilot allocation, which approximates the optimal allocation, outperforms equal allocation. However, as the number of bins increases, the total number of pilot samples also increases, resulting in wasted samples. Appendix \ref{appendix:ss_classifier_exps} contains additional results for stratification with different classifiers. In general, while simpler classifiers (see Appendix \ref{appendix:ss_classifier_exps}) are effective, Roberta dramatically reduces the number of samples needed to estimate the prevalence.

\begin{figure*}[ht]
    \centering
    \includegraphics[scale=0.94]{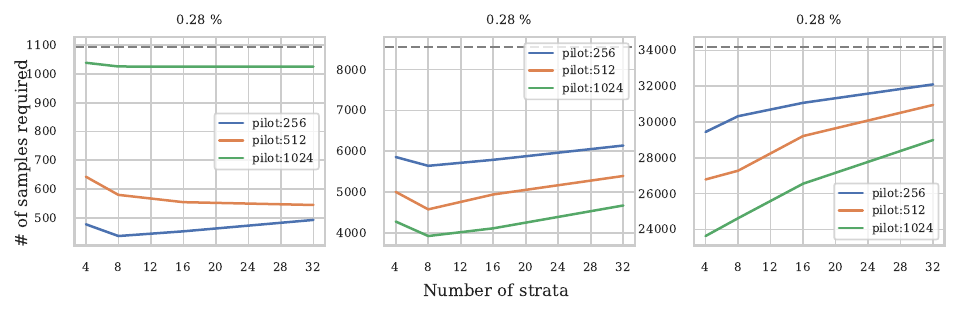}
    \caption{Number of samples required to estimate the prevalence of toxicity within 20\%, for different prevalences. All pilot results are averaged over 100 trials. The numbers associated with the pilot is the \textit{total} number of pilot samples across all bins. By choosing appropriate powers of two, each bin always has an integer number of pilot samples. Dotted lined represents random sampling. For small prevalences, more pilot samples are better, because more samples estimate the standard deviations within each bin better. In contrast, more bins fares worse because each bin has less pilot samples.}
    \label{fig:prevalence_binning}
\end{figure*}

\paragraph{Pilot sampling under different prevalences} We examine the robustness of pilot sampling to different class imbalances in Figure \ref{fig:prevalence_binning}. For each example, civilcomments has a real valued toxicity label corresponding to the proportion of annotators that labelled it as toxic. When converting the labels to binary, a threshold of 0.5 yields a prevalence of 5.9\%, which is the setting studied in Figure \ref{fig:binning_allocation}. In Figure \ref{fig:prevalence_binning} we present results for different prevalences by varying the threshold among $\{0.48, 0.83, 0.95\}$. Stratified sampling provides significant efficiency over random sampling. The number of pilot samples matters for both very small and very large prevalences. In the case of high prevalence, pilot samples outnumber the optimal allocation so there will be wasted samples. In the case of low prevalence, additional pilot samples help estimate the standard deviation of the strata correctly for optimal allocation. Finally, increasing the number of strata does not help with lower prevalences, as fewer samples are available for each bin to estimate the standard deviation with.

\paragraph{Recommendations} Based on our simulation in CivilComments, our recommendations for conducting stratified sampling in practice are:
\textbf{(1)} Choose quantile binning. Among the two practical binning methods, quantile binning outperforms equal width binning. It performs at near oracle binning performance if the number of bins is sufficiently large and the optimal allocation is used.
\textbf{(2)} Choose pilot allocation. Attempting to estimate a good allocation with pilot samples will greatly reduce the variance of the stratified sampling estimator.
\textbf{(3)} Make a reasonable choice for the number of pilot samples. If the prevalence is suspected to be low, or the estimation requirement is precise, inefficiency in pilot samples is less of a concern (because the pilot samples are needed anyways). When the prevalence is high and the number of samples required may be less than the total pilot samples, a reasonable choice needs to be made about the number of pilot samples collected. The experimenter must consider how much data annotation is acceptable at worst.

\section{Recall reporting in practice}
\label{sec:reddit_prevalence}

In this section, we present a case study of recall reporting for Reddit. As the requirement currently underspecifies the technical implementation, we identify a number of ways recall could be reported. We then choose one to report and discuss the implications and areas for legal clarification. 
We choose to focus on Reddit for two reasons: first, Reddit qualifies as an ``online platform'' as defined in Art. 3(i) DSA  and therefore is obliged to provide transparency reports annually (Art. 15(1) DSA).\footnote{\url{https://ec.europa.eu/commission/presscorner/detail/en/ip_23_2413}} Second, Reddit data is most readily available for academic study.

\subsection{Addressing underspecification}
In implementing a technical evaluation to satisfy ``accuracy'' reporting for the DSA, three basic specifications are needed:

\paragraph{Evaluation metric} As discussed in \S \ref{sec:dsa}, we establish recall as a legally sensible metric, which we showed in \S \ref{sec:simulated_results} can be efficiently estimated with stratified sampling.

\paragraph{Ground truth}
The DSA addresses two classes of content (see \S \ref{sec:content_moderation}). To report recall, we must consider how the true positives and true negatives can be annotated:
\begin{itemize}
    \item \emph{Illegal content.} 
    To identify illegal content, legal definitions  (e.g., punishable hate speech) would need to be operationalized as clear annotation guidelines \cite{zufall_hate_speech}. To be feasible at the scale social media requires, such identification likely needs to be possible for a skilled person with no legal training. 
    
    \item \emph{Content that violates terms and conditions.}
    A convenience in identifying whether speech violates community guidelines as a part of terms and conditions is that it is already operationalized by the platform. The measurement of accuracy on the guidelines is a measurement of whether the moderation system is achieving its intended purpose. This would also complement transparency measures about the guidelines themselves (see Art. 14 DSA). 
\end{itemize}
As the scope of the DSA applies to all content, and all platforms have community guidelines \cite{gillespie2018custodians}, this specification may be general enough to allow for explicit legal clarification. However, regulators should take caution when specifying recall on illegal content. Such reporting is only feasible if the respective legal bases defining content illegality are amenable to operationalization.

\paragraph{Test set} Finally, the set of content (or test set) to report the accuracy metric over needs to be decided. Since the DSA specifies bi-annual or annual reporting periods, it would be natural to evaluate on content accumulated within the last reporting period. The DSA does not provide further guidance, and we speculate two appropriate test sets:
\begin{itemize}
    \item \emph{All visible content.} An accuracy metric could be evaluated over all visible content on the platform, which would also follow industry best practices on how prevalence metrics are reported (see \S \ref{sec:related_work}).

    \item \emph{All notified content.} Art. 16 DSA requires providers to put in place mechanisms to allow users to notify (i.e., flag) content they consider to be illegal. When a notice is submitted, providers are liable if they do not act expeditiously to provide a legal examination and remove or disable access to this content if it is illegal (Art. 16(3), Art. 6(1)(b) DSA). As platforms are expected to review this content, the moderation recall of this subset may be especially important. However, anyone can flag content without concrete reason, possibly even with malicious intent \cite{meyer_2014}.
\end{itemize}
The specification of a test set will be heavily platform dependent. For Reddit, content may mean posts or comments, but for Youtube it may mean videos. In Snapchat, most content is not publicly visible, and it may be more appropriate to calculate recall over views. Regulators should be wary of overgeneralizing, as social media platforms have diverse functionality \cite{gillespie_expanding_2020}. Underspecifying the test set allows for custom implementation and may encourage meaningful reporting.

\paragraph{Our study} We study the moderation recall of subreddits, in terms of content removal as the means of moderation.\footnote{Content removal is the primary means of moderation here, see \url{https://support.reddithelp.com/hc/en-us/articles/15484284113172-What-mods-can-do}} For the purposes of our study, we choose to report the recall of personal attacks on all visible comments. We find this to reasonably satisfy the requirement as Reddit has community guidelines against such speech and most of the discussion happens in the comments. Practically, the other options require data or annotations that are not readily available for academic study. In our preliminary analysis of comments, we found that personal attacks can be identified consistently. This is corroborated by \citet{habernal-etal-2018-name}, which reports that personal attacks in r/changemyview have high inter-annotator agreement.

\subsection{Experimental setup}

\begin{table*}
    \centering
    \begin{tabular}{l|cc|cc|c|c|cc}
          & & & \multicolumn{2}{c|}{$n$ 
          for $\textsc{CV}_{\hat{p}_{st}} \leq 20\%$} & Reduction & Comments & Mod. & \\
          & $\hat{p}_{st}$ & 95\% CI & Stratified & Random & w/ strat. & removed & recall & 95\% CI \\
         \hline
         r/politics & 7.00\% & [6.42\%, 7.58\%] & 1033 & 1401 & 26\% & 0.77\% & 9.91\% & [9.22\%, 10.71\%] \\
         r/AskReddit & 4.75\% & [4.29\%, 5.21\%] & 1425 & 2142 & 33\% & 0.70\% & 12.92\% & [11.91\%, 14.11\%]  \\
         r/sex & 3.00\% & [2.63\%, 3.37\%] & 2477 & 3556 & 30\% & 2.06\% & 40.72\% & [37.94\%, 43.93\%] \\
         r/pcmasterrace & 4.00\% & [3.61\%, 4.39\%] &  1488 &  2567 & 42\% & 0.89\% & 18.20\% &  [16.86\%, 19.78\%] \\
         r/wow & 4.25\% & [3.76\%, 4.74\%] & 2078 & 2455 & 15\% & 0.75\% & 14.97\% & [13.63\%, 16.60\%] \\
         r/legaladvice & 1.50\% & [1.25\%, 1.75\%] & 5027 & 7591 & 33\% & 7.65\% & 83.62\% & [81.38\%, 85.97\%]\\
         \hline
    \end{tabular}
    \caption{Statistics derived from pilot annotations. 
    Stratified sampling was applied with 8 quantile bins, 50 pilot annotations per bin, and a Roberta finetuned on personal attack data. Estimating the prevalence here is equivalent to using equal proportion allocation.
    The number of samples required and efficiency for estimation with random and pilot stratified sampling is provided assuming the lower bound of prevalence.}
    \label{tab:pilot_statistics}
\end{table*}


 \paragraph{Study of interest} We restrict our study to a recent dump of Reddit (December 2022) on Pushshift.io \cite{DBLP:conf/icwsm/BaumgartnerZKSB20}. We select 6 subreddits with a range of macro-norm violation prevalence, as estimated in \citet{park_measuring_2022}, and that have more than 1000 comments per day\footnote{Accessed January 21st, 2023, as per \url{https://subredditstats.com/}.}:
\begin{quote}
    (most macro-norm violations) r/politics, r/AskReddit, r/sex, r/pcmasterrace, r/wow, r/legaladvice (least)
\end{quote}
At the time of writing, 4 out of the 6 subreddits we study explicitly have rules against ``personal attacks" (politics: rule 4, AskReddit: rule 8, wow: rule 1, legaladvice: rule 5). The other two subreddits: r/sex, r/pcmasterrace, also have rules regarding constructive engagement and respectful conversation, respectively. 

\paragraph{Data collection} To download the comments, we first randomly sample a post submission (akin to a thread) from the subreddit in the Pushshift data dump. We then use Praw\footnote{\url{https://praw.readthedocs.io/en/stable/}} to collect all the non-top level comments in the submission. We repeat this process until we have either 100K comments or have exhausted all possible comments for that subreddit.

\paragraph{Removed comments} While publicly viewable posts from Reddit are accessible through the Pushshift data stores, records for removed comments are incomplete. Therefore,  we can only derive an upper bound of content moderation recall by assuming every removal decision is a true positive. 

\paragraph{Personal attacks} Using the data from \citet{habernal-etal-2018-name}, we finetuned a Roberta binning classifier on a balanced training set with an equal number of personal attacks and negatives from r/changemyview. We label comments as personal attacks according to \citet{habernal-etal-2018-name} and additional annotation details are provided in Appendix \ref{appendix:annotation_details}.

\subsection{Discussion}
\label{sec:pe_results}

\paragraph{Results} Refer to Table \ref{tab:pilot_statistics}. 
As a sanity check, we see that ranking subreddits by personal attack prevalence aligns closely with their ranking by macro-norm violation as estimated in \citet{park_measuring_2022}. By the prevalences, we can conclusively determine that r/politics has the biggest problem with personal attacks out of all the subreddits we examine. 
By moderation recall, r/politics is also low. While they have a rule against personal attacks, we speculate that moderators avoid removing too many comments to appear politically neutral. For r/legaladvice, personal attack prevalence is already low, and moderators also remove many comments that are off topic, which may inflate the true positives and consequently the recall. In terms of methodology, applying our version of stratified sampling often yields substantial sample reductions. These gains are present despite distribution shift, where the binning classifier was trained on r/changemyview but tested on other subreddits.

\paragraph{Confidence intervals} Even though we collected a small number of samples, the confidence intervals on the recall are narrow because the relative difference between the true positives and false negatives was often large. This means that as the number of personal attacks removed surpasses the prevalence, the recall becomes easier to estimate. Since we were able to provide tight bounds around the recall with a small number of annotator hours, we expect it feasible for Reddit to report more comprehensive recall measures annually. 




\paragraph{Discrepancies with the reporting obligation} There are two discrepancies between our experiment and the actual reporting obligation of the DSA. First, our recall is reported over a month while Reddit only needs to provide transparency reports every year. Our method can easily be adopted to a larger pool and the estimation difficulty only increases in the larger pool if the prevalence decreases. Second, the removed comments we consider could have been removed through either automated or manual means, while only recall reporting of automated tools are required by the DSA. Since we do not have records of how the comments were removed, we cannot easily provide an isolated accuracy of the automated systems, but our method is general. However, we point out that the recall presented is in line with current industry prevalence metrics, and we additionally suggest legislators could expand the scope of the requirement to an aggregate moderation accuracy.

\section{Conclusion}
\label{sec:conclusion}


In this work, we identified ``accuracy'' in the DSA as an open-textured term, which requires further interpretation. To operationalize the requirement, we considered the regulatory purposes standing behind the DSA and the underlying fundamental rights which are affected. The DSA balances freedom of expression with the right to non-discrimination, and we find precision and recall to be an appropriate instantiation of these rights, where balancing is reflected in the precision-recall tradeoff. We derive an efficient and robust estimator, which we use to compute a moderation recall for Reddit. By providing a legally and technically sound interpretation of ``accuracy'', we highlight the implications and areas for further legal clarification. 
In light of the insights here, we hope regulators can provide further legal or regulatory clarification to prevent enforcement deficits and better fulfill the regulatory purpose of the DSA.

\section{Ethical considerations}

\paragraph{Risks of content moderation}
Content moderation can violate the right to the freedom of expression. Automated content moderation further raises ethical concerns about problematic training data or application of such systems perpetuating societal power dynamics \cite{gillespie_scale_2020}. We do not approve the use of these systems for censorship and have especially highlighted that automated content moderation needs to respect freedom of speech.
We promote the use of stratified sampling: it involves humans and is unbiased with respect to a human defined ground truth. Finally, our work focuses on the \emph{evaluation} of content moderation rather than directly making removal decisions, and we hope it can help further the study of fair interventions to reduce social media harms. 

\paragraph{Risks of regulatory capture} Regulation can be co-opted to serve minority interests. Facebook, for instance, has openly requested for additional social media regulation, which is speculated to serve their interest in preventing competition \cite{Macias_2020}. Our position on regulation is not neutral either, as this work is rooted in existing regulation. However, the entirety of this work is dedicated to content moderation recall's legal and technical sensibility. Our proposals introduce a meaningful transparency requirement which introduces strictly more reporting burden. A large portion of this work is dedicated to technical feasibility, which is especially important in considering the compliance of new companies and preventing regulatory capture.

\paragraph{Data statement} Our experiments are based on data sets that include postings qualifying as personal data in the sense of the EU General Data Protection Regulation. Even though the data sets have been made publicly available before, our experiments qualify as ``processing" and thus needs to be justified. We base this justification on Art. 6(1)(f), Art. 89 GDPR for purposes of scientific research.

\section*{Acknowledgements}
Discussion with Sagnik Ghosh of the Integrity Institute inspired the scope of our problem. 
Swabha Swayamdipta and the USC NLP group provided feedback on early versions of this work.
Over several conference submissions, many anonymous reviewers helped improve the quality and clarity of this work.
We thank all who have made our work possible.
This work was funded by grants from Open Philanthropy, Cisco Research, and Google Research.
\bibliography{aaai24}

\appendix

\section{Additional statistical background}

\subsection{Formulating prevalence estimation} \label{appendix:formulation}

To estimate $FN$, we refactor this estimation as a problem of estimating the prevalence $p$, or rate of false negatives among the predicted negatives. 
With the probability $p$, we could use $FN = p | \mathcal{N}|$ to obtain the number of false negatives, where  $\mathcal{N}$ are the predicted negatives.
Statistically, we estimate the false negative rate, or prevalence as:
\begin{equation} \label{expectation}
    p = \mathbb{E}_{x \sim \mathcal{N}}[s(x)]
\end{equation}
where $s(x) = \mathbb{I}(x \in S)$ is an binary function indicating whether the content $x$ is a true positive (i.e. violating or illegal content). 

The simplest way to estimate this probability is by applying the random sampling estimator $\hat{p}$ where 
\begin{equation}
    \hat{p} = \frac{1}{n} \sum^{n}_{i=1} s(x_i)
\end{equation}
and $\{ x_i, i=1...n \}$ is a random sample from the remaining content $\mathcal{U}$. This estimator has standard error
    $\textsc{SE}_{\hat{p}} = \sqrt{\hat{p}(1-\hat{p})/n}$,
which is directly proportional to width of the confidence interval on $p$ if a normal approximation is applied. This  quantifies how precisely we have estimated $p$.

\begin{table}[t]
    \centering
    \begin{tabular}{r|r|r|r}
         \diagbox{$p$}{Prec.} & 20\% & 10\% & 5\% \\
         \hline
         0.1 & 865 & 3458 & 13830 \\
         \textbf{0.059} & 1532 & 6127 & 24508 \\
         0.01 & 9508 & 38031 & 152122 \\
         0.001 & 95941 & 383762 & 1535047 \\
    \end{tabular}
    \caption{Number of samples needed for the random sampling estimator to estimate different prevalences, to different levels of precision. A normal approximation is assumed. The smaller the prevalence the more difficult it is to estimate precisely. Highlighted in bold is the prevalence of positives in the CivilComments dataset.}
    \label{tab:rs_power}
\end{table}

\subsection{Stratified sampling notation} \label{appendix:ss_notation}

We adopt notation from \citet{pss}.
Let $L$ be the number of strata, $N$ be the total size of the dataset, $N_h$ be the number of total items in stratum $h$, and $n_h$ be the number of samples drawn and annotated from stratum $h$ for estimation.
The stratified sampling estimator $\hat{p}_{st}$ is defined by the equations:
\begin{align}
    \hat{p}_{st} &= \sum_{h=1}^{L} \frac{N_h}{N} \cdot \hat{p}_h \\
    \text{where} ~~
    \hat{p}_{h} &= \frac1{n_h} \sum_{n_h}^{i=1} s(x^h_i) \, \forall \, h \in \{1, \dotsc, L\};
\end{align}
and $\{x^h_i\}_{i=1}^{n_h}$ is a random sample of $n_h$ examples from stratum $h$.
In other words, $\hat{p}_{st}$ is a weighted average of the $\hat{p}_h$'s, which are our prevalence estimates within each stratum. This estimator is unbiased, i.e., the expected value $\mathbb{E}[\hat{p}_{st}]$ of the estimator equals the true probability $p$. 

Given a stratification, we can analytically calculate the variance of the stratified estimator as
\begin{equation}
    \widehat{\text{Var}}(\hat{p}_{st}) = \sum_{h=1}^{L} \left(\frac{N_h}{N}\right)^2 \left(1 - \frac{n_h}{N_h}\right)^2 \cdot \frac{\hat{p}_h (1 - \hat{p}_h)}{n_h}.
\end{equation}
Note that the second term $(1 - n_h / N_h)$ is a finite population correction and $\approx 1$ when the strata are large relative to the number of annotations. The standard error is then $\textsc{SE}_{\hat{p}_{st}} = \sqrt{\widehat{\text{Var}}(\hat{p}_{st})}$ and the confidence intervals of the estimator can be calculated by assuming the estimator follows a normal distribution (by the central limit theorem).

\subsection{Optimal allocation for stratified sampling} \label{appendix:optimal_alloc}

For a given stratification, there is a closed form solution of an optimal allocation which minimizes the variance of the stratified sampling estimator. For each stratum, the number of samples to allocate is given by
\begin{equation} \label{optimal_allocation}
    n^{opt}_h = n \cdot \frac{N_h\sigma_h}{\sum_k N_k\sigma_k}
\end{equation}
where $\sigma_h$ is the standard deviation of samples within stratum $h$, and $n$ is the total number of planned samples to annotate. While these standard deviations cannot be known beforehand, an estimate can be made for $n^{opt}_h$ using a small pilot sample from each stratum. 

\subsection{Power calculations for random sampling}
\label{appendix:rs_power}

For random sampling we have
\begin{align}
    \textsc{SE}_{req}^2 = \frac{\hat{p}(1-\hat{p})}{n} \qquad
    \therefore n = \frac{\hat{p}(1-\hat{p})}{\textsc{SE}_{req}^2}
\end{align}
to achieve the desired accuracy.

For reference, the number of samples to estimate different prevalences, to different levels of precision, with random sampling is listed in Table \ref{tab:rs_power}.

\subsection{Power calculations for stratified sampling}
\label{appendix:cost_derivations}

Given a stratification, we can analytically calculate costs as well. For equal allocation stratified sampling, we have
\begin{align}
    \textsc{SE}_{req}^2 &= \sum_{h=1}^{L} \left(\frac{N_h}{N}\right)^2 \cdot \frac{\hat{p}_h (1 - \hat{p}_h)}{(1/L)n} \\
    \therefore n &= \left(\frac{L}{\textsc{SE}_{req}^2}\right) \sum_{h=1}^{L} \left(\frac{N_h}{N}\right)^2 \cdot \hat{p}_h (1 - \hat{p}_h)
\end{align}
where we simplified the variance term of the stratified estimator by dropping the population correction (if $N_h$ is large relative to $n$, the correction is approximately 1). For optimal allocation stratified sampling, we have
\begin{align}
    \textsc{SE}_{req}^2 &= \sum_{h=1}^{L} \left(\frac{N_h}{N}\right)^2 \cdot \frac{\hat{p}_h (1 - \hat{p}_h)}{n^{opt}_h} \\
    \therefore n &= \left(\frac{1}{\textsc{SE}_{req}^2}\right) \sum_{h=1}^{L} \left(\frac{N_h}{N}\right)^2 \cdot \frac{\hat{p}_h (1 - \hat{p}_h)}{c^{opt}_h}
\end{align}
where $c^{opt}_h$ is the proportion of $n$ allocated to stratum $h$.

\section{Estimating precision}
\label{appendix:estimating_precision}

To estimate precision, the true positives (i.e., violating or illegal content) need to be determined within the positives (i.e., removed content).
Statistically, we are estimating the false positive rate $q = \mathbb{E}_{x \sim \mathcal{P}}[s(x)]$
where $\mathcal{P}$ denotes the set of positives and $s(x) = \mathbb{I}(x \in S)$ is an binary function indicating whether the content $x$ is a true positive. The value of $q$ is then exactly the precision. $q$ is only difficult to estimate when it is almost 0 or 1. The former is unlikely, as companies won't deploy useless systems, and the latter would mean the system works well and exact precision estimates are unnecessary. 
We can construct a sample mean estimator $\hat{q}$, by sampling $N$ items from $\mathcal{P}$ and computing the percentage of correct removals. The distribution of this sample mean $\hat{q}$ is approximately normal, and reasonable confidence intervals are easy to obtain. 

The qualitative difficulty is in implementing the scoring function $s(x)$. Since the ground truth will be applied to content detected by an automated system, many false positives may be borderline. In the case where the ground truth is illegal content, a trained legal expert may be needed, and some borderline content may even need to wait for judicial decisions. In the case where the ground truth are community guidelines, the decision boundaries of these guidelines are continually refined, and the moderators judging content violation need substantial training \cite{newton_2019}.

\begin{table*}[]
    \centering
    \begin{tabular}{l|ll|rrr}
         &  &  & 20\% & 10\% & 5\% \\
         & Binning & Allocation & [28.9\%, 37.9\%] & [30.8\%, 35.2\%] & [31.2\%, 34.0\%] \\
         \hline
         Random sampling & n/a & n/a & 2247 & 8986 & 35942 \\
         \hline
         Unigram & oracle:8 & optimal & 991 & 3961 & 15844 \\
         & quantiles:8 & pilot:50 & $1242 \pm 45$ & $4965 \pm 176$ & $19936 \pm 784$ \\
         \hline
         Roberta (balanced) & oracle:8 & optimal & 353 & 1410 & 5639  \\
         & quantile:8 & pilot:50 & $940 \pm 39$  & $3758 \pm 153$ & $15025 \pm 627$ \\
         \hline
         Distilbert (balanced) & oracle:8 & optimal & 628 & 2511 & 10044  \\
         & quantile:8 & pilot:50 & $961 \pm 38$  & $3846 \pm 164$ & $15362 \pm 642$ \\
    \end{tabular}
    \caption{Number of samples needed for the stratified sampling estimator to estimate prevalence of toxicity (4.1\%) in the unfiltered CivilComments within 5, 10, and 20\%. The intervals denote the confidence intervals on the recall. Oracle:8 means 8 strata were created and pilot:50 means 50 pilot samples are taken within each strata for variance estimation. Oracle binning and optimal allocation are oracle techniques which rely on labelled data that is not available beforehand in practice. We chose 8 bins and 50 pilot samples because these settings were effective in Figure \ref{fig:binning_allocation}.}
    \label{tab:classifier_power}
\end{table*}

\section{Additional stratified sampling experiments}
\subsection{Stratification with different classifiers} \label{appendix:ss_classifier_exps}

We experiment using stratified sampling with a few supervised classification techniques. These classifiers are used to score the unlabeled examples, from which the strata are created.

\paragraph{Models} We study three models: a unigram logistic regression model, Roberta \cite{DBLP:journals/corr/roberta}, and Distilbert. Roberta \cite{DBLP:journals/corr/roberta} is a transformer-based language model which is pretrained on a large corpus of English data. Distilbert is smaller version of Roberta with the same model architecture but also 40\% less parameters. We find both pretrained models to have strong performance for civilcomments after finetuning. We use typical hyperparameters for small datasets \cite{DBLP:conf/iclr/MosbachAK21}, and make two adjustments that we found to be critical: increasing the batch size to 32 (we hypothesize that since the labels can be noisy, the gradients are unstable), and subsampling the negative examples so that the training data is balanced (the resulting training data has 11k examples).

\paragraph{Results} In Table \ref{tab:classifier_power}, we show that using a classifier can greatly reduce annotation requirements over a random sampling estimator. 
Estimating prevalence within 20\% of 4.1\% requires only a few hundred samples. While the unigram model is effective, Roberta dramatically reduces the number of samples needed to estimate the prevalence. The gap between quantile and oracle binning methods is larger for Roberta than for unigram models, suggesting that Roberta would benefit from a better binning method.

\section{Additional annotation details}
\label{appendix:annotation_details}

We label a comment as a personal attack if it fits any of criteria defined in \cite[Table 2]{habernal-etal-2018-name} including: vulgar insult, illiteracy insult, ridicule or sarcasm, condescension, accusation of stupidity, accusation of ignorance, and accusation of argumentative fallacies. 
We found most personal attacks to be immediately recognizable because it included insults or accusations. For borderline cases, such as rude comments, we labelled it a personal attack if it didn't engage with the parent comment. In some other cases, viewing the original comment thread of Reddit was helpful. When the comment was part of a ``flame war'', we would annotate it as a personal attack. All annotation was conducted by trusted and compensated annotator, and the rate of annotation was about 600 examples per hour. In a small interannotator agreement study with the first author, the agreement was substantial, with a Cohen's $\kappa$ of 0.69 (out of 100 examples, both annotators agreed on 96).

\end{document}